\documentclass[a4paper,aps,superscriptaddress,floatfix,nofootinbib,twocolumn,bibtex]{revtex4}

\usepackage{bbold}
\usepackage{bbm}
\usepackage[pdftex]{graphicx}
\usepackage{latexsym,amsmath,verbatim}
\usepackage{color}
\usepackage{rotating}
\usepackage{multirow}
\usepackage[english]{babel}

\begin{document}

\title{Abrupt transition
in the structural formation of interconnected networks}

\author{Filippo Radicchi}\email{f.radicchi@gmail.com}
\affiliation{Departament d'Enginyeria Quimica, Universitat Rovira i Virgili, 43007 Tarragona, Spain}
\author{Alex Arenas}
\affiliation{Departament d'Enginyeria Inform\'atica i Matem\'atiques, Universitat Rovira i Virgili, 43007 Tarragona, Spain}


\begin{abstract}
\noindent  Our current world is linked by a complex mesh of networks where information, people and goods flow. These networks are interdependent each other, and present structural and dynamical features
different from those observed in isolated networks.
While examples of such ``dissimilar'' properties are becoming more abundant, for example diffusion, robustness and competition, it is not yet clear where these differences are rooted in. Here we show that the composition of independent networks into an interconnected network of networks undergoes a structurally sharp transition as the interconnections are formed. Depending of the relative importance of inter and intra-layer connections, we find that the entire interdependent system can be tuned between two regimes: in one regime, the various layers are structurally decoupled and they act as independent entities; in the other regime, network layers are indistinguishable and the whole system behave as a single-level network. We analytically show that the transition between the two regimes is discontinuous even for finite size networks. Thus, any real-world interconnected system is potentially at risk of abrupt changes in its structure that may reflect in new dynamical properties.
\end{abstract}


\maketitle

\noindent Interacting, interdependent or multiplex networks are different ways of naming the same class of complex systems where networks are not considered as isolated entities but interacting each other. In multiplex, the nodes at each network are instances of the same entity, thus the networks are representing simply different categorical relationships between entities, and usually categories are represented by layers. Interdependent networks is a more general framework where nodes can be different at each network.
\\
Many, if not all, real networks are ``coupled'' with other real networks. Examples can be found in several domains: social networks (e.g., Facebook, Twitter, etc.) are coupled because they share the same actors~\cite{lamb}; multimodal transportation networks are composed of different layers (e.g., bus, subway, etc.) that share the same locations~\cite{barth}; the functioning of communication and power grid systems depend one on the other~\cite{buldy}. So far, all phenomena that have been studied on interdependent networks, including percolation~\cite{buldy, grass}, epidemics~\cite{mendiola}, and linear dynamical systems~\cite{gomez}, have provided results that differ much from those valid in the case of isolated complex networks. Sometimes the difference is radical: for example, while isolated scale-free networks are robust against failures of their nodes or edges~\cite{jeong}, scale-free interdependent networks are instead very fragile~\cite{buldy, grass}.
\\
Given such observations, two fundamentally important theoretical questions are in order: 
(i) Why do dynamical and critical phenomena running on interdependent network models differ so much from their analogous in isolated networks?;
(ii) What are the regimes of applicability of the theory valid for isolated networks to interdependent networks? In this paper, we  provide an analytic answer to both these questions by characterizing the structural properties of the whole interconnected network in terms of the networks that compose it.

\begin{figure}
\begin{center}
\includegraphics[width=0.45\textwidth]{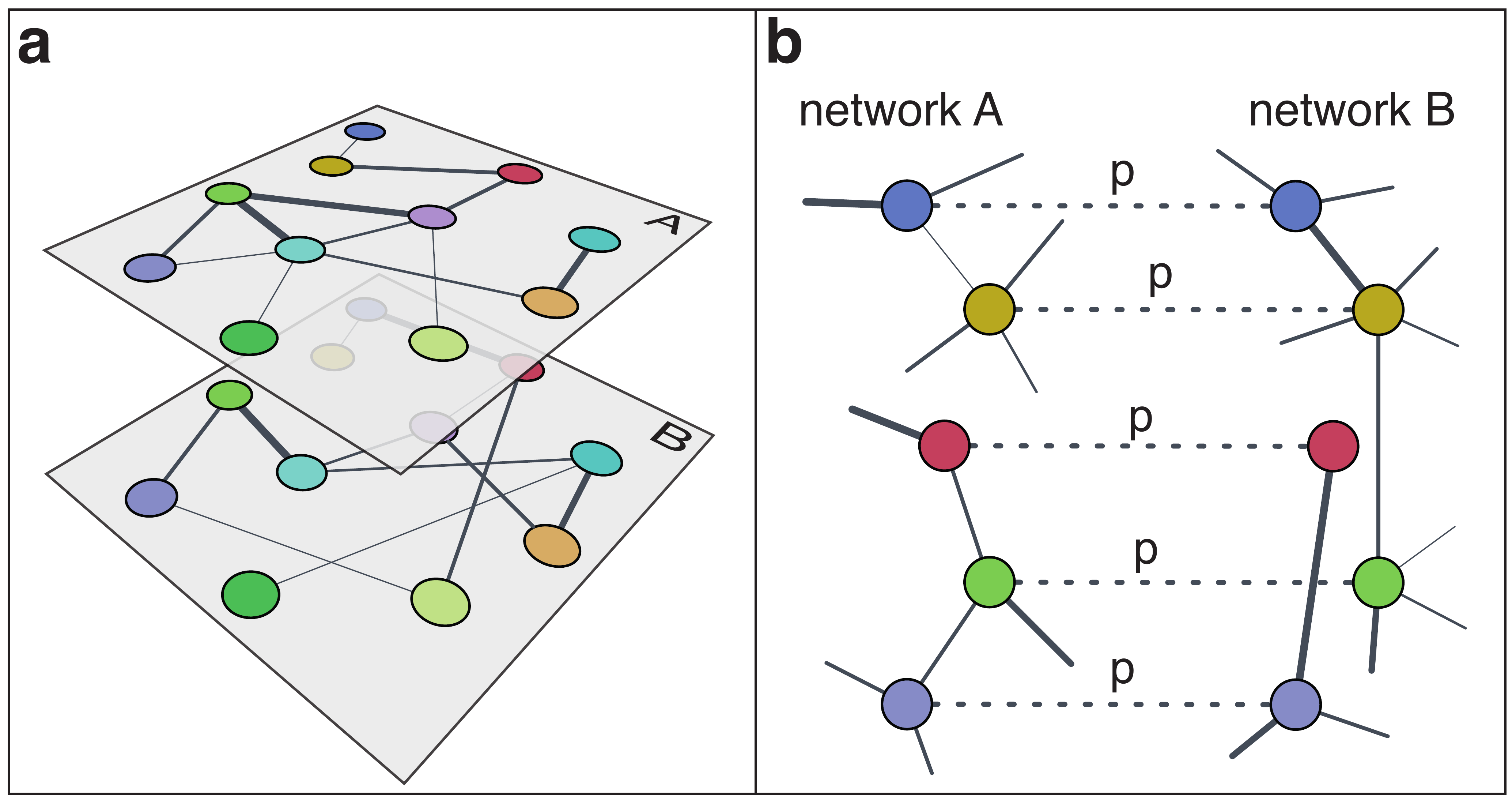}
\end{center}
\caption{{\bf a)} Schematic example of two interdependent networks $A$ and $B$. In this representation, nodes
of the same color are one-to-one interdependent. {\bf b)} 
In our model, inter-layer edges have  weights equal to $p$.
}
\label{fig:1}
\end{figure}

\

\noindent For simplicity, we consider here the case of two interdependent networks. The following method can be, however, generalized to an arbitrary number of interdependent networks and its solution is reported in the Supplementary Information.
We assume that the two interdependent networks $A$ and $B$ are undirected and weighted, and that they have the same number of nodes $N$.
The  weighted adjacency matrices of the two graphs are indicated as $A$ and $B$, respectively, and they have both dimensions $N \times N$. With this notation, the element $A_{ij} = A_{ji}$ is equal to the weight of the connection between the nodes $i$ and $j$ in network $A$. The definition of $B$ is analogous.
\\
We consider the case of one-to-one symmetric interdependency~\cite{buldy} between nodes in the networks $A$ and $B$ (see Fig.~1A). In the more
general case of multiple interdependencies, the solution
is analogous and reported in the Supplementary Information.
The connections between interdependent nodes of the two networks are weighted by a factor $p$ (see Fig.~1B), any other weighted factor for the networks $A$ and $B$ is implicitly absorbed in their weights.
The supra-adjacency matrix $G$ of the whole network is therefore given by
\begin{equation}
G = \left(
\begin{array}{cc}
A  & p \mathbbm{1}\\
p \mathbbm{1} & B
\end{array} 
\right) \, ,
\label{eq:adj}
\end{equation}
where $\mathbbm{1}$ is the identity matrix of dimensions $N \times N$.
\\
Using this notation we can define the supra-laplacian of the interconnected network as
\begin{equation}
\mathcal{L} = 
 \left(
\begin{array}{cc}
\mathcal{L}_{A} + p \mathbbm{1} & - p \mathbbm{1}\\
- p \mathbbm{1} & \mathcal{L}_{B} + p \mathbbm{1}
\end{array} 
\right) \, .
\label{eq:lap}
\end{equation}
The blocks present in $\mathcal{L}$ are square symmetric matrices of dimensions $N \times N$,
In particular, $\mathcal{L}_{A}$ and $\mathcal{L}_{B}$ are the laplacians of the networks $A$ and $B$, respectively.
\\
Our investigation focus on the analysis of the spectrum of the supra-Laplacian to ascertain the origin of the structural changes of the merging of networks in an interconnected system. The spectrum of the laplacian of a graph is a fundamental mathematical object for the study of the structural properties of the graph itself. There are many applications and results on graph Laplacian eigenpairs and their relations to numerous graph invariants (including connectivity, expanding properties, genus, diameter, mean distance, and chromatic number) as well as to partition problems (graph bisection, connectivity and separation, isoperimetric numbers, maximum cut, clustering, graph partition), and approximations for optimization problems on graphs (cutwidth, bandwidth, min-p-sum problems, 
ranking, scaling, quadratic assignment problem)~\cite{merris, chungbook, eigbook, chung}.
\\
Note that, for any graph, all eigenvalues of its laplacian are non negative numbers. The smallest
eigenvalue is always equal to zero and the eigenvector associated to it is trivially a vector whose entries are all identical.
The second smallest eigenvalue $\lambda_2$ also called the {\em algebraic connectivity}~\cite{fiedler1} is one of
the most significant eigenvalues of the Laplacian. It is strictly larger than zero only if the graph is connected.
More importantly, the eigenvector associated to $\lambda_2$, which is called the {\em characteristic valuation} 
or {\em Fiedler vector} of a graph, provides even deeper about its structure~\cite{fiedler2, fiedler3, mohar}. For example, the components of this vector associated to the various nodes of the network are used in spectral clustering algorithms
for the bisection of graphs~\cite{clustering}.

\

\noindent Our approach consists in the study of the behavior of the second smallest eigenvalue
of the supra-laplacian matrix $\mathcal{L}$ and its characteristic valuation as a function of $p$,
given the single-layer network laplacians $\mathcal{L}_A$
and $\mathcal{L}_B$.
\\
According to the theorem by Courant and Fisher (i.e., the so-called
min-max principle)~\cite{courant, fisher}, 
the second smallest eigenvalue of $\mathcal{L}$ is given by
\begin{equation}
\lambda_2\left(\mathcal{L}\right) = \min_{\left|v\right> \in \mathcal{V}} \,  \left< v \right| \mathcal{L} \left| v \right> \; ,
\label{eq:minmax}
\end{equation}
where
$\left| v \right> \in  \mathcal{V} \, \textrm{ is such that } \,  \left< v|1\right> =0 , \left< v|v\right> =1$.
\\
The vector $\left|1\right>$ has $2N$ entries all equal to $1$.
Eq.~(\ref{eq:minmax}) means that
$\lambda_2\left(\mathcal{L}\right)$ is 
equal to the minimum of the
function $ \left< v \right| \mathcal{L} \left| v \right>$, over all possible
vectors $\left|v\right>$ that are orthogonal to
the vector $\left|1\right>$ and that
have norm equal to one.
The vector for which such minimum is reached is thus
the characteristic valuation
of the supra-laplacian
(i.e., $\mathcal{L} \left|v\right> = \lambda_2 \left|v\right>$).
\\
We distinguish two blocks of size $N$ in the vector $\left|v\right>$ by writing it as 
$\left|v\right>= \left|v_A, v_B\right>$. In this notation, $\left|v_A\right>$ is the part of the
eigenvector whose components corresponds to the nodes of network $A$, while 
$\left|v_B\right>$ is the part of the eigenvector whose components corresponds to the
nodes of network $B$. We can now write
\[
\begin{array}{ll}
\left< v \right| \mathcal{L} \left| v \right> = & \left< v_A, v_B \right| \mathcal{L} \left| v_A, v_B \right> = 
\\
& \left<v_A\right|\mathcal{L}_A\left|v_A \right> + \left<v_B\right|\mathcal{L}_B\left|v_B \right>  + \\
& p \left( \left<v_A | v_A \right>   
+ \left<v_B | v_B \right> - 2 \left<v_A|v_B\right> \right) 
\end{array}
\]
and the previous set of constraints as  
$\left< v_A|1\right> + \left< v_B|1\right> =0$ and $\left< v_A|v_A\right> + \left< v_B|v_B\right> =1$,  
where now all vectors have dimension $N$.
Accounting for such constraints, we can finally rewrite the minimization problem as
\begin{equation}
\begin{array}{l}
\lambda_2\left(\mathcal{L}\right) = p +  \min_{ \left|v\right> \in \mathcal{V}} \, \left\{  \left<v_A\right|\mathcal{L}_A\left|v_A \right>  \right.
\\ \left.  +  \left<v_B\right| \mathcal{L}_B\left|v_B \right> - 2 p \left<v_A|v_B\right> \right\}
\end{array} \; .
\label{eq:eig}
\end{equation}
This minimization problem can be solved using Lagrange
multipliers (see Supplementary Information for technical details).
\\
In this way we are able to find that the second smallest eigenvalue of the supra-laplacian
matrix $\mathcal{L}$ is given by
\begin{equation}
\lambda_2\left(\mathcal{L}\right) =
\left\{
\begin{array}{ll}
2p & \textrm{ , if } p\leq p^*\\
\leq \frac{1}{2} \lambda_2\left( \mathcal{L}_A + \mathcal{L}_B\right)  & 
\textrm{ , if } p\geq p^*
\end{array}
\right. \; .
\label{eq:final}
\end{equation}
Thus indicating that the algebraic connectivity of the interconnected system follows two distinct
regimes, one in which its value is independent of the structure of the two layers, and the other
in which its upper bound is limited by the algebraic connectivity of the weighted superposition of the two layers
whose laplacian is given by $ \frac{1}{2} \left( \mathcal{L}_A +  \mathcal{L}_B\right)$.
More importantly, the discontinuity in the first derivative of $\lambda_2$ is reflected in a radical change
of the structural properties of the system happening at $p^*$ (see Supplementary Information). Such dramatic change
is visible in the coordinates of characteristic valuation of the nodes of the two network layers. In the regime $p\leq p^*$, the components of the
eigenvector are 
\begin{equation}
\left|v_A\right> = 
- \left|v_B\right> \qquad \textrm{ where } \left|v_A\right> = \pm \frac{1}{\sqrt{2N}} \left|1\right> \; .
\label{eq:sign1}
\end{equation}
This means that the two network layers are structurally disconnected and independent. For $p\geq p^*$, we have
\begin{equation}
 \left<v_A|1\right>=\left<v_B|1\right>=0  \; ,
\label{eq:sign2a}
\end{equation}

\begin{figure*}
\begin{center}
\includegraphics[width=0.95\textwidth]{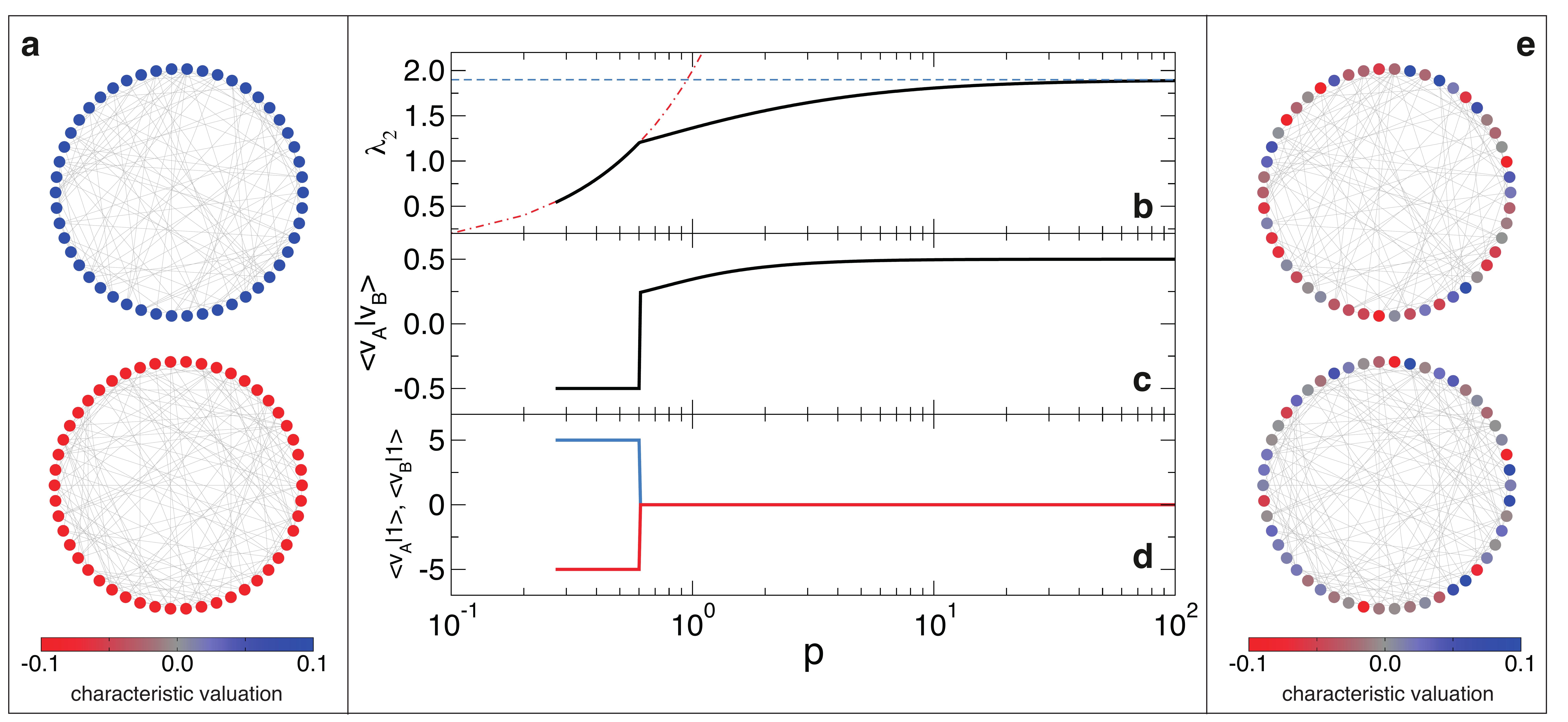}
\end{center}
\caption{Algebraic connectivity and
Fiedler vector for two interdependent
Erd\H{o}s-R\'enyi networks
of $N=50$ nodes and
average degree $\bar{k}=5$. In this example, the critical
point is $p^*=0.602(1)$. 
{\bf a)} Characteristic valuation of the nodes
in the two network layers for $p=0.602$.
{\bf b)} Algebraic connectivity of the system (black line).
The discontinuity of the first derivative
of $\lambda_2$ is very clear. 
The two different regimes $2p$ and 
$\frac{\lambda_2\left(\mathcal{L}_A+\mathcal{L}_B\right)}{2}$
are shown as red dot-dashed and blue dashed lines, respectively.
{\bf c)} Inner product
$\left<v_A|v_B\right>$ between the
part of the Fiedler eigenvector ($\left|v_A\right>$)
corresponding to nodes in the network $A$
and the one ($\left|v_B\right>$) corresponding to vertices
in network $B$ as a function of
$p$. {\bf d)} Inner products $\left<v_A|1\right>$ and 
$\left<v_B|1\right>$ as functions of
$p$. $\left<v_A|1\right>$ and $\left<v_B|1\right>$
indicate the sum of all components
of the Fiedler vectors $\left|v_A\right>$ and $\left|v_B\right>$, 
respectively.
{\bf e)} Characteristic valuation of the nodes
in the two network layers for $p=0.603$.
}
\label{fig:2}
\end{figure*}

\noindent which means that the components of the vector corresponding to interdependent
nodes of network $A$ and $B$ have the same sign, while nodes in the same layer have
alternating signs. Thus in this second regime, the system connectivity is dominated by inter-layer connections, and
the two network layers are structurally indistinguishable.
\\
The critical value $p^*$ at which the transition occurs is the point at which we observe the
crossing between the two different behaviors of $\lambda_2$, which means
\begin{equation}
p^* \leq \frac{1}{4}\, \lambda_2\left( \mathcal{L}_A + \mathcal{L}_B\right) \; .
\label{eq:critic}
\end{equation}
This upper bound becomes exact in the case of identical network layers (see Supplementary Information).
Since inter-layer connections have weights that grows with $p$, the transition happens at the point at which
the weight of the inter-layer connections exceeds the half part of the inverse of the algebraic connectivity of the
weighted super-position of both network layers (see Fig.~2). In the case of $\ell$ network layers, the result is equivalent 
to the superposition of all of them (see Supplementary Information). 
\\
It is important to notice that the discontinuity in the first derivative of $\lambda_2\left(\mathcal{L}\right)$ can be 
interpreted as the consequence of the crossing of two different populations of eigenvalues (see the case of identical 
layers in the Supplementary Information).
The same crossing will also happen for the other eigenpairs of the graph laplacian (except for the
smallest and the largest ones), and thus
will reflect in the discontinuities in the first derivatives of 
the corresponding eigenvalues.

\noindent A physical interpretation of the algebraic phase transition that we are able to analytically 
predict can be given by viewing the function $\left<v\right|\mathcal{L}\left|v\right>$ as an energy-like function. From this point of view, 
Eq.~(\ref{eq:minmax}) becomes equivalent to a search for the ground state energy, and the characteristic valuation can be viewed as the ground state configuration. 
Such analogy is straightforward if one realizes that Eq.~(\ref{eq:minmax}) is equivalent to the minimization of the weighted cut
of the entire networked system [whose adjacency matrix $G$ is defined in Eq.~(\ref{eq:adj})], and that the minimum of this function corresponds to the
ground state of a wide class of energy functions~\cite{kolmo} and fitness landscapes~\cite{land}.
These include, among others, the energy associated to the Ising spin models~\cite{parisi} and costs functions of combinatorial optimization problems,
such as the traveling salesman problem~\cite{sale}. In summary, the structural transition of interdependent networks involves a discontinuity in the first
derivative of an energy-like function, and thus, according to the Ehrenfest classification of phase transitions, 
it is a discontinuous transition~\cite{phase}.
\\
Since the transition at the algebraic level has the same nature as the connectivity transition
that has been studied by Buldyrev {\it et al.} in the same class of networked systems~\cite{buldy},
it is worth to discuss about the relations between the two phase transitions.
We can reduce our model to the annealed version of the model considered by  Buldyrev {\it et al.}
by setting $A=t^2 A$, $B=t^2 B$ and $p=t$, being $1-t$ the probability that one node
in one of the networks fails. All the results stated so far hold, with only two different interpretations.
First, the upper bound of Eq.~(\ref{eq:critic}) becomes a lower bound for the critical threshold of the algebraic transition
that reads in terms of occupation probability as
\begin{equation}
t_c \geq \frac{4}{\lambda_2\left(\mathcal{L}_A + \mathcal{L}_B\right)} 
\; .
\label{eq:perc}
\end{equation}
Second, the way to look at the transition must be reversed: network layers are structurally independent (i.e., the analogous 
of the non percolating phase) for values of $t\leq t_c$, while become algebraically connected (i.e., analogous of the percolating phase)
when $t \geq t_c$. 
\\
As it is well known, the algebraic connectivity represents a lower bound for both the edge connectivity and node connectivity of
graph (i.e., respectively the minimal number of edges or nodes that should be removed to disconnect the graph)~\cite{fiedler1}.
Indeed, the algebraic connectivity  of a graph is often used as a control parameter to make the graph more resilient to random failures
of its nodes or edges~\cite{jama}.
Thus, the lower bound of Eq.~(\ref{eq:perc}) represents also a lower bound for the critical percolation
threshold measured by Buldyrev {\it et al}. Interestingly, our prediction turns out to be a sharp estimate of the lower bound.
\\
For the Erd\H{o}s-R\'enyi model,  we have in fact  $t_c \geq 2/ \bar{k}$, if the two networks
have the same average degree $\bar{k}$, and this value must be compared with
$2.455/\bar{k}$ as predicted by Buldyrev {\it et al.}~\cite{buldy, grass}.
Similarly, we are able to predict  that $t_c$ grows as the degree distribution
of the network becomes more broad~\cite{chung}, in the same way as it has been numerically
observed by Buldyrev {\it et al.}~\cite{buldy}.
\\
Although we are not able to directly map the algebraic transition to the percolation
one, we believe that the existence of a first-order transition at the
algebraic level represents an indirect support of the discontinuity of the percolation
transition. 

\

\noindent In conclusion, we have provided the exact analytic treatment of the structural properties of interconnected networks.
We have presented the exact solution for the algebraic connectivity of these network models.
For simplicity, we have considered the simplest case of one-to-one interdependency
but our formalism can be easily extended to study more complicated dependence relationships
among the nodes of the different layers. Our proof does not rely on any approximation 
but on a very intuitive mathematical approach. 
\\
The  structural phase transitions in interdependent networks are first-order in nature.
This differentiate multi- and single-level networks in a radical manner. We remark that the discontinuity
in the first derivative of the algebraic connectivity affects directly a vast class of
systems  whose dynamics is driven by the minimization of energy-like functions
associated to the structure of the system, but the same conclusions can be also extended to other critical phenomena
whose features depend on the third, fourth, etc. smallest eigenpairs of the graph laplacian.
\\
Moreover, the point at which we observe the discontinuity in the
first derivative of the algebraic connectivity (but also on other eigenvalues of the graph laplacian) defines a clear scale
for the applicability of the results valid for isolated networks. In one case, network layers can be
considered as independent, in the other case the entire system can be considered as
a single-level network. The fact that the transition between the two regimes is so sharp leaves out only a very tiny 
interval of interaction values where it makes sense to consider the system as composed of many interacting network layers. 
\\
Our results have also deep practical implications. The abrupt
nature of the structural transition is not only visible in the limit of infinitely large systems, but
for networks of any size. Thus, even real networked systems composed of few elements
may be subjected to abrupt structural changes, including failures. Our theory provides, however, 
fundamental aids for the prevention of such collapses. It allows, in fact, not only
the prediction of the critical point of the transition, but, more importantly, to
accurately design the structure of such systems in order to make them more robust. For example, the
percolation threshold of interconnected systems can be simply decreased by increasing
the algebraic connectivity of the superposition of the network
layers. This means that an effective strategy to make an interdependent system more
robust is to avoid the repetition of edges among layers, and thus bring the
superposition of the layers as close as possible to an all-to-all topology.

 \begin{acknowledgments}
\noindent This work has been partially supported by the
Spanish DGICYT Grants FIS2012-38266, FET projects PLEXMATH 
(318132) and the Generalitat de Catalunya 
2009-SGR-838.
F.R. acknowledges
support from the Spanish Ministerio de Ciencia e Innovaci´on
through the Ram\'on y Cajal program.
A.A. acknowledges the ICREA Academia
and the James S. McDonnell Foundation.
\end{acknowledgments}

\newpage

\renewcommand{\theequation}{S\arabic{equation}}
\setcounter{equation}{0}
\renewcommand{\thefigure}{S\arabic{figure}}
\setcounter{figure}{0}
\renewcommand{\thetable}{S\arabic{table}}
\setcounter{table}{0}

\section*{Supplementary Information}

\subsection*{Solution of the algebraic connectivity value of interconnected networks}

\noindent In the following, we will make use of the standard bra-ket notation 
for vectors. In this notation, $\left|x\right>$ indicates a column vector, $\left<x\right|$ indicates the transposed
(i.e., row vector) of $\left|x\right>$, $\left<x|y\right>=\left<y|x\right>$ indicates the inner product between
the vectors $\left|x\right>$ and $\left|y\right>$, $A\left|x\right>$ indicates the action of 
matrix $A$ on the column vector $\left|x\right>$, and $\left<x\right|A$ indicates the action of 
matrix $A$ on the row vector $\left<x\right|$.

\

\noindent First of all, we can simply state that for the algebraic connectivity of Eq.~(\ref{eq:eig}) we must have that
\begin{equation}
\lambda_2\left(\mathcal{L}\right) \leq  \frac{1}{2} \lambda_2\left( \mathcal{L}_A+  \mathcal{L}_B\right) \; ,
\label{eq:bound}
\end{equation}
where this upper bound comes out directly from the definition
of the minimum of a function. For every $\mathcal{Q} \subseteq \mathcal{V}$, we have in fact that
\[
 \min_{ \left|v\right> \in \mathcal{V}} \left<v\right|\mathcal{L}\left|v\right>  \leq  \min_{ \left|v\right> \in \mathcal{Q}} \left<v\right|\mathcal{L}\left|v\right> 
\]
simply because we are restricting the domain
in which finding the minimum of the function
$\left<v\right|\mathcal{L}\left|v\right> $.
The particular value of the upper bound
of Eq.~(\ref{eq:bound}) is then given by setting
$\mathcal{Q}$ as
\[
\begin{array}{l}
\left| v \right> = \left| v_A, v_B \right> \in  \mathcal{Q} \, \textrm{ is such that } \, \left|v_A\right> = \left|v_B\right> = \left|q\right> 
\\
\textrm{ , with } \left< q|1\right> =0, \left< q|q\right> = 1/2 
\end{array}
\;.
\] 
\\
To find the minimum of the function
expressed in Eq.~(\ref{eq:eig}), we
use the Lagrange multipliers' formalism. This means finding the minimum of the function
\[
\begin{array}{ll}
M = &  \left<v_A\right| \mathcal{L}_A\left|v_A \right>  +   \left<v_B\right|\mathcal{L}_B\left|v_B \right>  
- 2 p \left<v_A|v_B\right>  
\\ & - r \left(\left< v_A|1\right> + \left< v_B|1\right>\right) - s \left(\left< v_A|v_A\right> + \left< v_B|v_B\right> - 1 \right)
\end{array}
\; ,
\]
where the constraints of the minimization problem have been explicitly inserted in the function
to minimize through the Lagrange multipliers $r$ and $s$. In the following calculations, 
we will make use of the identities
\[
\begin{array}{l}
\frac{\partial \,}{\partial\, \left|x\right>} \left<t|x\right> = \frac{\partial \,}{\partial\, \left|x\right>} \left<x|t\right> =  \left<t\right|
\\
\frac{\partial \,}{\partial\, \left|x\right>} \left<x|x\right> =  2\left<x\right| 
\\  
\frac{\partial \,}{\partial\, \left|x\right>} \left<x\right|A\left|x\right>=
2\left<x\right|A \textrm{, if } A=A^T 
\end{array} 
\; ,
\]
where $\frac{\partial \,}{\partial\, \left|x\right>}$ indicates the derivative with respect to all the coordinates of the vector $\left|x\right>$.
Equating to zero the derivatives of $M$ with respect to $r$ and $s$, we obtain the constraints that we imposed. 
By equating to zero the derivative of $M$ with respect to $\left|v_A\right>$, we obtain instead
\begin{equation}
\frac{\partial \, M}{\partial\, \left|v_A\right>} = 2  \left<v_A\right|\mathcal{L}_A - 2 p \left<v_B\right| - r \left<1\right| -2 s \left< v_A \right| = \left<0\right| \; ,
\label{der1}
\end{equation}
and, similarly for the derivative of $M$ with respect to $\left|v_B\right>$,we obtain
\begin{equation}
\frac{\partial \, M}{\partial\, \left|v_B\right>} = 2  \left<v_B\right|\mathcal{L}_B - 2 p \left<v_A\right| - r \left<1\right| -2 s \left< v_B \right| = \left<0\right| \; .
\label{der2}
\end{equation}
Multiplying both equations for $\left|1\right>$, we have
$2  \left<v_A\right|\mathcal{L}_A\left|1\right> - 2 p \left<v_B|1\right> - r \left<1|1\right> -2 s \left< v_A|1\right> =0$ and
$2  \left<v_B\right|\mathcal{L}_B\left|1\right> - 2 p \left<v_A|1\right> - r \left<1|1\right> -2 s \left< v_B|1\right> =0$, that can be simplified
in $2 (p-s) \left<v_A|1\right> - r N =0$ and
$2 (p-s) \left<v_B|1\right> - r N  =0$ 
because $\mathcal{L}_A\left|1\right> = \mathcal{L}_B\left|1\right> = \left|0\right>$ and $\left<v_A|1\right> = - \left<v_B|1\right>$. 
Summing them, we obtain  $r =0$.
Finally, we  can write
\begin{equation}
\begin{array}{l}
(p-s) \left<v_A|1\right> = 0\\
(p-s) \left<v_B|1\right>  = 0
\end{array} \; .
\label{eq:s}
\end{equation}
These equations can be true in two cases: (i) 
$\left<v_A|1\right> \neq 0$ or $\left<v_B|1\right>\neq0$
and $s=p$; (ii) 
$\left<v_A|1\right>=\left<v_B|1\right>=0$. In the following,
we analyze these two cases separately.

\

\noindent First, let us suppose that
$s=p$, and that at least one of the
two equations $\left<v_A|1\right>\neq 0$
and $\left<v_B|1\right>\neq 0$ is true. If
we set $s=p$ in Eqs.~(\ref{der1}) 
and~(\ref{der2}), they become
\begin{equation}
  \left<v_A\right|\mathcal{L}_A - p \left<v_B\right| - p \left< v_A \right| = \left< 0 \right|
\label{eq:tt1}
\end{equation}
and 
\begin{equation}
  \left<v_B\right|\mathcal{L}_B - p \left<v_A\right| - p \left< v_B \right| = \left< 0 \right| \; .
\label{eq:tt2}
\end{equation}
If we multiply the first equation for $\left|v_A\right>$
and the second equation for $\left|v_B\right>$, 
the sum of these two new equations is
\begin{equation}
 \left<v_A\right|\mathcal{L}_A\left|v_A\right> +  \left<v_B\right|\mathcal{L}_B\left|v_B\right> - 2 p \left<v_A|v_B\right>  = p  \; .
\label{eq:condition}
\end{equation}
If we finally insert this expression in Eq.~(\ref{eq:eig}),
we find that the second smallest eigenvalue
of the supra-laplacian is
\begin{equation}
\lambda_2\left(\mathcal{L}\right) = 2 p \; .
\label{eq:sol1}
\end{equation}
We can further determine the components of
Fiedler vector in this regime. If we take the 
difference between Eqs.~(\ref{eq:tt1})
and~(\ref{eq:tt2}), we 
have $ \left<v_A\right|\mathcal{L}_A =  \left<v_B\right|\mathcal{L}_B$.
On the other hand,
Eq.~(\ref{eq:sol1})
is telling us that 
 $ \left<v_A\right|\mathcal{L}_A\left|v_A\right> = 
- \left<v_B\right|\mathcal{L}_B\left|v_B\right>$
because the only term surviving
in Eq.~(\ref{eq:condition}) is the 
one that depends on $p$.
Since $\left<v_A\right|\mathcal{L}_A\left|v_A\right>$
($\left<v_B\right|\mathcal{L}_B\left|v_B\right>$) is always
larger than zero, unless $\left|v_A\right>= c \left|1\right>$
($\left|v_B\right>= c \left|1\right>$), with $c$ arbitrary constant value,
we obtain Eq.~(\ref{eq:sign1}).
Thus in this regime, both the relations
$\left<v_A|1\right> \neq 0$ 
and $\left<v_B|1\right> \neq 0$ must be simultaneously true.
Eq.~(\ref{eq:sign1}) 
also means that $\left<v_A|v_B\right> = - \frac{1}{2}$. 
\\


\

\noindent The other possibility is that Eqs.~(\ref{eq:s}) are satisfied
because $\left<v_A|1\right> =0 $ 
and $ \left<v_B|1\right> = 0$ are simultaneously
true. 
In this case, the average value of the components
of the vectors $\left|v_A\right>$ and
$\left|v_B\right>$ is zero, and thus
the coordinates of the Fiedler vector
corresponding to the nodes
of the same layer have alternatively negative
and positive signs.
More can be said in the case of identical layers,
where the problem can be solved exactly (see next section).
In this case, the upper bound of Eq.~(\ref{eq:bound}) 
becomes the exact solution for the algebraic
connectivity and reads as
$\lambda_2 \left(\mathcal{L}\right) = 
\lambda_2 \left(\mathcal{M}\right)$,
with $\mathcal{M}$ laplacian of both layers.
More importantly, the Fiedler vector satisfies 
the relation
\begin{equation}
\left|v_A\right> = \left|v_B\right> \; .
\label{eq:sign2}
\end{equation}
The same relation does not hold in general
for different network layers, although
the coordinates of the Fiedler
vector of two interdependent
nodes  seem to have the same sign.

\subsection*{Spectrum of the laplacian
for two identical network layers}

\noindent Consider the case $\mathcal{L}_A=\mathcal{L}_B=\mathcal{M}$.
Finding the eigenvalues of the supra-laplacian
$\mathcal{L}$ means finding the
solutions of 
the eigenvalue problem 
\[
\textrm{det} \left(\mathcal{L} - \lambda \mathbbm{1}\right) =0 \; .
\]

\noindent Let us write the eigenvalues $\lambda$ as functions
of the eigenvalues $\mu$ of $\mathcal{M}$. This
can be done in the following way.

\[
\left(\mathcal{L} - \lambda \mathbbm{1}\right) = 
\left(
\begin{array}{cc}
 \mathcal{M} + p \mathbbm{1} - \lambda \mathbbm{1}  & - p \mathbbm{1}\\
- p \mathbbm{1}  &  \mathcal{M} + p \mathbbm{1} -  \lambda \mathbbm{1}
\end{array}
\right)
\]

\noindent Consider the matrices

\[
U = 
\left(
\begin{array}{cc}
Q & \emptyset \\
\emptyset & Q
\end{array}
\right)
\]
\[
U^{T} = 
\left(
\begin{array}{cc}
Q^{T} & \emptyset \\
\emptyset & Q^{T}
\end{array}
\right) \; ,
\]

\noindent with $Q^{T} \mathcal{M} Q = D$ and $D$ diagonal
matrix containing the eigenvalues
$\mu$ of $\mathcal{M}$, so that
$Q^{T} Q = Q Q^{T} = \mathbbm{1}$, and the matrices

\[
V = \frac{1}{\sqrt{2}} \left(
\begin{array}{cc}
\mathbbm{1} & -\mathbbm{1}\\
\mathbbm{1} & \mathbbm{1}
\end{array}
\right)
\]

\[
V^{T} = \frac{1}{\sqrt{2}} \left(
\begin{array}{cc}
\mathbbm{1} & \mathbbm{1}\\
- \mathbbm{1} & \mathbbm{1}
\end{array}
\right)
\]

We can write
\[
\begin{array}{l}
V^{T} U^{T} \left(\mathcal{L} - \lambda \, \mathbbm{1} \right) U V =
\\
V^T
  \left(
\begin{array}{cc}
 D + p \mathbbm{1} - \lambda \mathbbm{1}  & - p \mathbbm{1}\\
- p \mathbbm{1}  &  D + p\mathbbm{1} -  \lambda \mathbbm{1}
\end{array}
\right)
V
\end{array}
\]

\[
\begin{array}{l}
V^{T} U^{T} \left(\mathcal{L} - \lambda \, \mathbbm{1} \right) U V =
\\
 \frac{1}{\sqrt{2}}\left(
\begin{array}{cc}
 D  - \lambda \mathbbm{1} &  D  - \lambda \mathbbm{1} \\
-  D  + \lambda \mathbbm{1} - 2 p \mathbbm{1} &  D  - \lambda \mathbbm{1} + 2 p  \mathbbm{1}
\end{array}
\right)
V
\end{array}
\]

\[
\begin{array}{l}
V^{T} U^{T} \left(\mathcal{L} - \lambda \, \mathbbm{1} \right) U V =
\\
  \frac{1}{2}\left(
\begin{array}{cc}
2 D  - 2 \lambda \mathbbm{1} & \emptyset \\
\emptyset & 2 D  - 2 \lambda \mathbbm{1} + 4 p \mathbbm{1}
\end{array}
\right)
\end{array} \; .
\]

\noindent Since
\[
\textrm{det} \left(\mathcal{L} - \lambda \mathbbm{1} \right) = \textrm{det} \left[ V^{T} U^{T} \left( \mathcal{L} - \lambda \mathbbm{1} \right) U V\right] \; ,
\]

\noindent the eigenvalues of the supra-laplacian
$\mathcal{L}$ are given by $\{ \mu\}$ and 
$\{ \mu + 2 p\}$, where $\left\{\mu\right\}$
are the eigenvalues of the single layer
laplacian $\mathcal{M}$.
\\
This means that there two possible candidates
for $\lambda_2\left(\mathcal{L}\right)$: 
$ \mu_2$
and $2 p$. The equation
that delimits the different
regions is thus
\[
 \mu_2\left(\mathcal{M}\right) = 2 p \; .
\]

\begin{figure*}
\begin{center}
\includegraphics[width=0.95\textwidth]{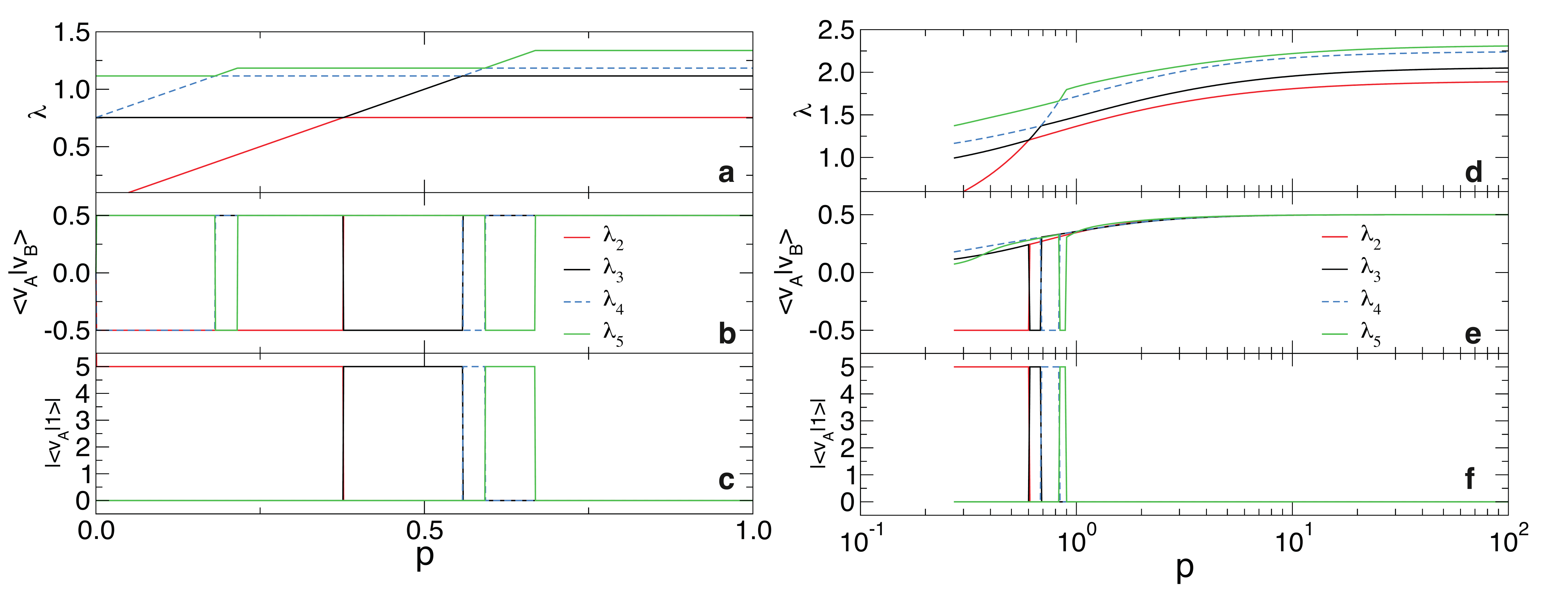}
\end{center}
\caption{Properties of some eigenpairs of the
supra-laplacian matrix
for two interdependent
Erd\H{o}s-R\'enyi networks
of $N=50$ nodes and
average degree $\bar{k}=5$. The networks
used in this plot are the same as those considered in Fig.~\ref{fig:2}.
In panels {\bf a}, {\bf b} and {\bf c}, we used identical layers
(only network $A$ for both layers), in panels 
{\bf d}, {\bf e} and {\bf f}, we used instead different
network layers.
{\bf a)} and {\bf d)} Eigenvalues $\lambda_k$, with $k=2, 3, 4$ and $5$
as functions of $p$.
{\bf b)} and {\bf e)} Inner product
$\left<v_A|v_B\right>$ between the
part of the eigenvector ($\left|v_A\right>$)
corresponding to nodes in the network $A$
and the one ($\left|v_B\right>$) corresponding to vertices
in network $B$ as a function of $p$.
{\bf c)} and {\bf f)} Absolute value of the inner product $\left<v_A|1\right>$
as a function of $p$.
}
\label{fig:3}
\end{figure*}

\noindent Please note that a similar behavior
is valid also for the other
eigenvalues of the laplacian
(except the largest and the smallest, see Fig.~\ref{fig:3}).
For example, the third smallest eigenvalue
$\lambda_3$ of the supra-laplacian
exhibits three different behaviors, an its derivative is 
discontinuous at two values of $p$ identified
by the equations
\[
 \mu_2\left(\mathcal{M}\right) = 2 p
\]
(i.e., the same point in which 
the first derivative of $\lambda_2$
is discontinuous) and
\[
 \mu_3\left(\mathcal{M}\right) =  \mu_{2}\left(\mathcal{M}\right)  + 2 p \; .
\]
The behavior of the other eigenvalues is even richer, 
and in principle several discontinuity points are
present. A similar behavior is also present
in the case of different network layers (see Fig.~\ref{fig:3}).


\subsection*{Spectrum
of the laplacian
with arbitrary number
of identical interconnected networks}

\noindent The same result holds also for
more than two identical interdependent networks. 
In that case, the matrix $V$ is the block matrix
able to diagonalize the block matrix composed of $\ell$
blocks equal to the identity matrix. $U$ is
still the matrix able to diagonalize
the laplacian $\mathcal{M}$. The resulting matrix, after
the similarity transformation
\[
V^{T} U^{T} \left( \mathcal{L} - \lambda \mathbbm{1} \right) U V
\]
has one block diagonal element
equal to 
$D + \ell \, p \mathbbm{1} - \lambda \mathbbm{1}$, and the remaining
$\ell-1$ block diagonal elements proportional
to $D - \lambda \mathbbm{1}$.
The eigenvalues of the supra-laplacian matrix are
thus $\{ \mu\}$ with multiplicity
$\ell-1$, and $\{ \mu + \ell p\}$
with multiplicity one. We thus have
still two regimes for the second smallest
eigenvalue given by
\[
\lambda_2\left(\mathcal{L}\right) =
\left\{
\begin{array}{ll}
\ell p & \textrm{ , if } p \leq p^*\\
 \mu_2\left(\mathcal{M}\right) &  \textrm{ , if } p \geq p^*
\end{array}
\right. \; ,
\]
where $p^*$ is given by
\[
p^* = \frac{1}{\ell} \mu_2\left(\mathcal{M}\right)\; .
\]

\subsection*{General case with arbitrary number
of interconnected networks}

\noindent Let us consider the case
of $\ell$ different  layers. The supra-laplacian
matrix is composed of $\ell \times \ell$
block matrices of dimensions $N \times N$. 
Along the diagonal,
we have
\[
\mathcal{L}_{mm} = \mathcal{L}_m + (\ell-1) p \mathbbm{1}
\]
while on the off-diagonal blocks we have
\[
\mathcal{L}_{mn} = - p \mathbbm{1} \; ,
\]
where $\mathcal{L}_{m}$ is the laplacian matrix of
the layer $m$, while $\mathbbm{1}$ is the
identity matrix.
Let us write the generic vector as
\[
\left|v \right> = \left|v_1, v_2, \ldots, v_\ell \right> \; .
\]

\noindent Then
\[
\begin{array}{l}
\left<v\right|\mathcal{L}\left|v \right> =
 \sum_m \left<v_m\right|\mathcal{L}_m\left|v_m\right> 
\\
+ (\ell-1) p \sum_m \left<v_m|v_m\right> - p \sum_m \sum_{n \neq m}  \left<v_m|v_n\right> 
\end{array} \; .
\]

\noindent For the Courant-Fisher min-max theorem, the second smallest
eigenvalue $\lambda_2\left(\mathcal{L}\right)$ of the
supra-laplacian matrix is given by
\[
0 \leq \lambda_2\left(\mathcal{L}\right) = \min_{\mathcal{V}} \left<v\right|\mathcal{L}\left|v \right> \; ,
\]
with 
\[
\begin{array}{l}
\left|v\right> \in  \mathcal{V} \textrm{ is such that } 
\\
\left|v\right> \neq \left|0\right>, 
\left<v|v\right>=1 \textrm{ and } \left<v|1\right> =0 
\end{array} \; .
\]
$\left|1\right>$ is the column vector 
whose $\ell N$ entries are equal to one,
while $\left|0\right>$ is the column 
vector whose $\ell N$ entries are equal to zero.
The constraints of the vectors in $\mathcal{V}$ can be written
also as
\[
\left<v|v\right>= \sum_m \left<v_m|v_m\right> =1 
\; \textrm{ and } \; \left<v|1\right>= \sum_m \left<v_m|1\right> =0 
\; ,
\]
where $\left|1\right>$ now indicates
a column vector whose $N$ entries are equal to one, and
$\left|0\right>$ now indicates
a column vector whose $N$ entries are equal to zero.
Imposing the constraint $\sum_m \left<v_m|v_m\right> =1$, the former expression
reduces to
\begin{equation}
\begin{array}{l}
\left<v\right|\mathcal{L}\left|v \right> =
\sum_m \left<v_m\right|\mathcal{L}_m\left|v_m\right>
\\
- p \sum_m \sum_{n \neq m}  \left<v_m|v_n\right> + (\ell-1) p 
\end{array} \; . 
\label{eq:eig_app}
\end{equation}

\noindent First of all, we can easily set an upper bound for 
$\lambda_2\left(\mathcal{L}\right)$ by
simply reducing the set of vectors where searching for the minimum of the function
$\left<v\right|\mathcal{L}\left|v \right>$.
For all $\mathcal{Q} \subseteq \mathcal{S}$, the definition
of minimum implies that
\[
\begin{array}{l}
\lambda_2\left(\mathcal{L}\right) \leq \min_{\mathcal{Q}} \, \left<v\right|\mathcal{L}\left|v \right> = (\ell-1) p 
\\
+ \, \min_{\mathcal{Q}} \,
\left[ \sum_m  \left<v_m\right|\mathcal{L}_m\left|v_m\right>  - p \sum_m \sum_{n \neq m}  \left<v_n|v_m\right> \right] 
\end{array} \; .
\]
In particular, if we choose $\mathcal{Q}$ 
\[
\begin{array}{l}
\left|v\right> = \left|v_1, \ldots, v_m\right> \in \mathcal{Q} \textrm{ is such that } \\
 \left|v_m\right> = \left|q\right> \textrm{ for all } m \textrm{ with } \left<q|1\right>  =0 \textrm{ and } \left<q|q\right>=1/\ell
\end{array} 
\]
this leads to
\[
\sum_m \sum_{n \neq m}  \left<v_n|v_m\right> = \sum_m \sum_{n \neq m}  \left<q|q\right> = \sum_m (\ell -1)/\ell  = \ell-1
\]
and therefore to
\[
\begin{array}{l}
\lambda_2\left(\mathcal{L}\right) \leq  \min_{\mathcal{Q}} \,
\sum_m \left<v_m\right|\mathcal{L}_m\left|v_m\right> = 
\\
\min_{\mathcal{Q}} \,  \left<q\right|\sum_m \, \mathcal{L}_m\left|q\right> = \frac{\lambda_2\left(\sum_m \mathcal{L}_m\right)}{\ell} \; .
\end{array}
\]
Notice that this upper bound does not depends
on $p$, and thus
represents
the asymptotic
value of $\lambda_2\left(\mathcal{L}\right)$ in the limit
$p \to \infty$. This can be proven in the following way. 
In the regime $p \gg 1$, 
we can write
\[
\begin{array}{l}
\min_{\mathcal{V}} \,   \left[ \sum_m \left<v_m\right|\mathcal{L}_m\left|v_m\right> - p \sum_m \sum_{n \neq m}  \left<v_m|v_n\right> \right] \sim \\
  \min_{\mathcal{V}_{p \gg 1}} \;  \left[ - p \sum_m \, \sum_{n \neq m}  \left<v_m|v_n\right>  \right] =  
\\
-p \; \max_{\mathcal{V}_{p \gg 1}} \;  \left[ \sum_m \, \sum_{n \neq m}  \left<v_m|v_n\right>  \right] 
\end{array}
\; .
\]
In this regime, the
terms $\left<v_m\right|\mathcal{L}_m\left|v_m\right>$ 
are in fact finite (i.e., they do not diverge with $p$), because
$\mathcal{L}_m$ does not depend on $p$ and because
the constraint $\sum_m \left<v_m|v_m\right> =1$ 
implies that $\left<v_m|v_m\right> \leq 1$. This basically means that
each component of the vector $\left|v_m\right>$ is in modulus smaller or 
equal to one. 
For the Cauchy-Swartz inequality, we can also write
\[
\left<v_n | v_m\right>^2 \leq \left<v_n | v_n\right> \, \left<v_m | v_m\right>
\]
and thus 
\[
\left<v_n | v_m\right> \leq  \sqrt{\left<v_n | v_n\right> \, \left<v_m | v_m\right>} \; .
\]
On the other hand, we have also that
\[
\begin{array}{l}
1 = \left(\sum_m \left<v_m | v_m\right> \right)^2 = 
\\
\sum_m \sum_{n \neq m} \left<v_n | v_n\right> \left<v_m | v_m\right> +  \sum_m \left<v_m | v_m\right>^2
\end{array}
\]
thus
\[
\begin{array}{l}
\sum_m \sum_{n \neq m} \left<v_n | v_n\right> \left<v_m | v_m\right> = 
\\
1 - \sum_m \left<v_m | v_m\right>^2 \leq 1
\end{array}
\]
This implies that
\[
\sum_m \sum_{n \neq m} \left<v_n | v_m\right> \leq 1
\]
where the equality holds only if 
all vectors $\left| v_m\right>$ are identical.
The maximum of the function thus corresponds
to one of these configurations, and thus
$\mathcal{V}_{p \gg 1} = \mathcal{Q}$.
This analytically prove the result
established by G\'omez {\it et al.}~\cite{gomez}
through approximation methods.

\

\noindent We can further investigate the structure
of the eigenvector associated to the eigenvalue
$\lambda_2\left(\mathcal{L}\right)$.
In order to find $\lambda_2\left(\mathcal{L}\right)$, we have to minimize
the function $\left<v\right|\mathcal{L}\left|v \right>$ under the constraints
of $\mathcal{V}$. This can be performed with the
use of the Lagrange multipliers, by minimizing the function
\[
\begin{array}{l}
M = \sum_m \left[ \left<v_m\right|\mathcal{L}_m\left|v_m\right>  - p \sum_{n \neq m}  \left<v_m|v_n\right> \right]  
\\
- r \left(\sum_m \left<v_m|v_m\right> - 1\right) - s \sum_m \left<v_m|1\right> 
\end{array} \; .
\]
By equating the derivatives of $M$ with respect to $r$ and $s$ we
simply recover the constraints. By equating to zero the
derivative of $M$ with respect to $\left|v_m\right>$, we
find
\begin{equation}
\begin{array}{l}
\frac{\partial \, M}{\partial \, \left|v_m\right>} = 
2 \left<v_m\right|\mathcal{L}_m \\ - 2p \sum_{n \neq m} \left<v_n\right| - 2r \left<v_m\right| - s \left< 1 \right| = \left< 0 \right|
\end{array}
\label{eq:der}
\end{equation}
where $\left<0\right|$ indicates
a row vector whose $N$ entries are equal to zero.
If we multiply the previous equation for $\left|1\right>$, we have
\[
2 \left<v_m\right|\mathcal{L}_m\left|1\right>  - 2p \sum_{n \neq m} \left<v_n|1\right> - 2r \left<v_m|1\right> - s \left< 1|1\right> = 0
\]
from which
\[
- 2p \sum_{n \neq m} \left<v_n|1\right> - 2r \left<v_m|1\right> - s N = 0
\]
because the $\mathcal{L}_m\left|1\right>=0$ and $\left<1|1\right>=N$. We further
have from one the constraints that 
$\sum_{n \neq m} \left<v_n|1\right> = - \left<v_m|1\right>$, thus
\begin{equation}
2\left(p-r\right) \left<v_m|1\right> - s N = 0 \; .
\label{eq:a} 
\end{equation}
If we sum the previous equation over all $m$, we have
\[
 2\left(p-r\right) \sum_m \left<v_m|1\right> - \sum_m s N = 0 \;  
\]
and since $\sum_m\, \left<v_m|1\right> =0$, we have
\[
s = 0 \;  .
\]
If we set $s=0$ in  Eq.~(\ref{eq:a}), we have
\[
\left(p - r\right) \left<v_m|1\right> = 0 \, , \; \forall \, m \; .
\]
These $\ell$ equations
are satisfied if: (i) $r = p$ 
and $\exists \, n $ such that $\left<v_n|1\right> \neq 0$, or
(ii) $\left<v_m|1\right> =0 $, $\forall \, m$.

\

\noindent Let us first suppose the first case, and thus 
$r = p$. Multiply Eq.~(\ref{eq:der}) for $\left|v_m\right>$ to obtain
\[
\left<v_m\right|\mathcal{L}_m\left|v_m\right>  -p \sum_{n \neq m} \left<v_n|v_m\right> - p \left<v_m|v_m\right>  = 0
\]
and summing over all layers $m$, we have
\[
\sum_m  \left[ \left<v_m\right|\mathcal{L}_m\left|v_m\right>  -p \sum_{n \neq m} \left<v_n|v_m\right> - p \left<v_m|v_m\right>  \right]  = 0 \; .
\]
If we now insert this expression in Eq.~(\ref{eq:eig_app}), we obtain
\[
\begin{array}{l}
\left<v\right|\mathcal{L}\left|v \right> =
\sum_m  \left[ \left<v_m\right|\mathcal{L}_m\left|v_m\right> -
 p \sum_{n \neq m}  \left<v_m|v_n\right> - \right.
\\
\left. p \left<v_m|v_m\right> + p \left<v_m|v_m\right>  \right]+ (\ell-1) p
\end{array}
\]
from which
\[
\left<v\right|\mathcal{L}\left|v \right> = p + (\ell -1) p = \ell p \; .
\]
Thus, in this regime, we have that
\[
\lambda_2\left(\mathcal{L} \right) = \ell p \; .
\]
\\
Since there is no dependency on $p$, we must have that
\[
\sum_m  \left<v_m\right|\mathcal{L}_m\left|v_m\right> =0  \; .
\]
This equation can be true
only if $\left|v_m\right> = c_m \left|1\right>$, with
$c_m$ arbitrary constant, and thus only if 
$\left<v_m\right|\mathcal{L}_m\left|v_m\right> =0$, $\forall \, m$.
This follows from the fact that 
$\left<x\right|\mathcal{L}_m\left|x\right> \geq 0$ for any
choice of $\left|x\right>$ and the equality holds
only for $\left|x\right> = c \left|1\right>$.
The relation between the constants $c_m$ is then
given by the normalization
\[
\sum_m \left<v_m|v_m\right> = N \sum_m c_m^2 = 1
\]
but also by the fact that
\[
\sum_m \left<v_m|1\right> = N \sum_m c_m = 0
\]
and there exists at least one $n$ for which
\[
c_n \neq 0 \; .
\]
In the case of $\ell=2$ layers, this reduces
to only one possibility as given by Eq.~(\ref{eq:sign1}).

\

\noindent  In conclusion, we can write that
\begin{equation}
\lambda_2\left(\mathcal{L} \right) = \min \, \left\{ \ell p, \mu_2\left(\mathcal{L}\right) \right\} \; ,
\label{eq:eig1}
\end{equation}
where 
\begin{equation}
\mu_2\left(\mathcal{L}\right) = \min_{\mathcal{T}} \left<v\right|\mathcal{L}\left|v \right> 
\label{eq:prob2}
\end{equation}
and
\[
\begin{array}{l}
\left|v\right> = \left|v_1, \ldots, v_m, \ldots, v_\ell\right> \in \mathcal{T} \textrm{ is such that }  \\
 \sum_m \left< v_m |v_m\right>=1 
\textrm{ and }  \left<v_m|1\right> =0 \,,  \forall\;  m 
\end{array} \;.
\]

\subsection*{Arbitrary interdependency matrix}

\noindent We consider here the case
$\ell=2$ network layers, but the calculations
are analogous for the case arbitrary $\ell$.
Suppose that the connections
between interdependent nodes in 
the networks $A$ and $B$ are described
by the symmetric matrix $C$. The supra-adjacency matrix is thus
\begin{equation}
G = \left(
\begin{array}{cc}
A  & p C \\
p C & B
\end{array} 
\right) \, ,
\label{eq:ar_adj}
\end{equation}
and the supra-laplacian matrix
is
\begin{equation}
\mathcal{L} = \left(
\begin{array}{cc}
\mathcal{L}_{A} + p D_C  & - p C \\
- p C & \mathcal{L}_{B} + p D_C
\end{array} 
\right) \, ,
\label{eq:ar_lap}
\end{equation}
where $D_C$ is the diagonal matrix
whose elements are $\left(D_C\right)_{ii} = \sum_j C_{ij}$.
We can write
\[
\begin{array}{l}
\left<v_A,v_B\right|\mathcal{L}\left|v_A,v_B\right> = \left<v_A\right|\mathcal{L}_{A}\left|v_A\right> + p \left<v_A\right|D_C\left|v_A\right> + 
\\
\left<v_B\right|\mathcal{L}_{B}\left|v_B\right> + p \left<v_B\right|D_C\left|v_B\right> - 2 p \left<v_A\right|C\left|v_B\right> \; .
\end{array}
\]
Proceeding in the same way as described
before (i.e., minimization with the use
of Lagrange multipliers), we
obtain the two following equations
\[
2 \left<v_A\right|\mathcal{L}_A + 2 p \left<v_A\right|D_C - 2 p \left<v_B\right|C - 2 s \left<v_A\right| - r \left<1\right| = \left<0\right|
\]
and
\[
2\left<v_B\right|\mathcal{L}_B + 2 p \left<v_B\right|D_C - 2 p \left<v_A\right|C - 2 s \left<v_B\right| - r \left<1\right| = \left<0\right| \; .
\]
If we multiply them for $\left|1\right>$, we have
\[
2 p \left<v_A|c\right> - 2 p \left<v_B|c\right> - 2 s \left<v_A|1\right> - r N = 0
\]
and
\[
2 p \left<v_B|c\right> - 2 p \left<v_A|c\right> - 2 s \left<v_B|1\right> - r N = 0 \; ,
\] 
where $\left|c\right> = C\left|1\right> = D_C\left|1\right>$ is
the vector whose coordinates correspond to the
strengths of the nodes in the interdependent part of the
graph. Summing them, we find $r=0$.
If we multiply the first equation for $\left|v_A\right>$, we have
$\left<v_A\right|\mathcal{L}_A\left|v_A\right> + p \left<v_A\right|D_C\left|v_A\right> - p \left<v_B\right|C\left|v_A\right> - s \left<v_A|v_A\right> = 0$
and $\left<v_B\right|\mathcal{L}_B\left|v_B\right> + p \left<v_B\right|D_C\left|v_B\right> - p \left<v_A\right|C\left|v_B\right> - s \left<v_B|v_B\right> = 0$,
thus from their sum we obtain
$s = \left<v_A, v_B\right|\mathcal{L}\left|v_A, v_B\right>$.
\\
If $C$ is the adjacency matrix
of a regular graph with degree $c$, then $\left|c\right> = c \left|1\right>$.
This means that
\[
 \left(2 pc - s\right) \left<v_A|1\right> = \left(2 pc - s\right) \left<v_B|1\right> = 0 \; .
\]
As in the former case, we can have two possibilities
\[
\left<v_A|1\right> = \left<v_B|1\right> = 0 
\]
or 
\[
\lambda_2\left(\mathcal{L}\right) = 2 pc \qquad \textrm{ with } \left<v_A|1\right> \neq 0 \;, \; \left<v_B|1\right> \neq 0\; .
\]

\section*{Annealed interconnected networks}

\noindent With the presented methodological approach, we can easily  study the typical behavior of different ensembles of network models. 
In this case, the adjacency matrices $A$ and $B$ should be thought as weighted  symmetric matrices where the weight of each edge
is equal to the probability of having a connection between nodes
in the ensemble of networks (i.e., 
so-called annealed networks~\cite{doro}). 
For example, if networks $A$ and $B$ are
Erd\H{o}s-R\'enyi models with connections probability $q_A$ and
$q_B$, respectively, the laplacian
of network $A$ is such that
$\left(\mathcal{L}_A\right)_{ij} = q_A (N-1)$ if $i=j$, and
$\left(\mathcal{L}_A\right)_{ij} = - q_A $, otherwise.
Similarly, we have
$\left(\mathcal{L}_B\right)_{ij} = q_B (N-1)$ if $i=j$, and
$\left(\mathcal{L}_B\right)_{ij} = - q_B $, otherwise.
The algebraic connectivity of $ \mathcal{L}_A +  \mathcal{L}_B$ can
be analytically estimated to be
$\lambda_2 \left( \mathcal{L}_A +  \mathcal{L}_B \right)=
\left( q_A +  q_B\right) N =  \bar{k}_A + 
 \bar{k}_B $, 
with $\bar{k}_A  = q_A N $ average degree of
network $A$ and $\bar{k}_B  = q_B N $
average degree of
network $B$.
Thus, the critical threshold of Eq.~(\ref{eq:critic}) becomes
$p^* \leq \left( \bar{k}_A  +  \bar{k}_B\right)/4$.
For more general network models, such annealed
networks with prescribed power-law degree 
distributions, the
critical point of the transition
can be also analytically
estimated by implementing the methodology
developed by Chung {\it et al.}~\cite{chung}.


\begin{thebibliography}{99}


\bibitem{buldy}
S. V. Buldyrev, R. Parshani, G. Paul, H. E. Stanley and S. Havlin. 
{\it Nature} {\bf 464}, 1025--1028 (2010).

\bibitem{gao}
J. Gao, S. V. Buldyrev, H. E. Stanley and S. Havlin.
{\it Nat. Phys.} {\bf 8}, 40--48 (2012). 

\bibitem{grass}
S. -W. Son, G. Bizhani, C. Christensen, P. Grassberger
and M. Paczuski. 
{\it EPL} {\bf 97}, 16006 (2012).


\bibitem{mendiola}
A. Saumell-Mendiola, M. \'A. Serrano and M. Bogu\~n\'a. 
{\it Phys. Rev. E} {\bf 86}, 026106 (2012).

\bibitem{gomez}
S. G\'omez, A. D\'iaz-Guilera, J. G\'omez-Garde\~nes, C.J. P\'erez-Vicente, Y. Moreno and A. Arenas.
{\it Phys. Rev. Lett.} {\bf 110}, 028701 (2013).

\bibitem{aguirre}
J. Aguirre, D. Papo and J. M. Buld\'u.
{\it Nat. Phys.} {\bf 9}, 230--234 (2013).


\bibitem{reka}
R. Albert and A. -L. Barab\'asi.
{\it Rev. Mod. Phys.} {\bf 74}, 47--97 (2002).

\bibitem{newmanbook}
M. E. J. Newman. 
{\it Networks: An Introduction.}
(Oxford University Press, New York, 2010).


\bibitem{doro}
S. N. Dorogovtsev, A. V. Goltsev and
J. F. F. Mendes. 
{\it Rev. Mod. Phys.} {\bf 80}, 1275--1335 (2008).


\bibitem{lamb}
M. Szella, R. Lambiotte and S. Thurner.
{\it Proc. Natl. Acad. Sci. USA} {\bf 107}, 13636--13641 (2010).


\bibitem{barth}
M. Barth\'elemy.
{\it Phys. Rep.} {\bf 499}, 1--101 (2011).



\bibitem{jeong}
R. Albert, H. Jeong and A.-L. Barab\'asi. 
{\it Nature} {\bf 406}, 378--382 (2000).


\bibitem{merris}
R. Merris. 
{\it Linear Algebra and its Applications} {\bf 197--198}, 143--176 (1994).

\bibitem{chung}
F. Chung, L. Lu and V. Vu.
{\it Proc. Natl. Acad. Sci. USA}  {\bf 100}, 6313--6318 (2003).

\bibitem{chungbook}
F. Chung.
{\it Spectral Graph Theory}. 
(CBMS Regional Conference Series in Mathematics, American Mathematical Society, 1997).

\bibitem{eigbook}
T. Biyikoglu, J. Leydold and P. F. Stadler. 
{\it Laplacian eigenvectors of graphs: Perron-Frobenius and Faber-Krahn type theorems}. (Lecture notes in mathematics, Springer-Verlag, Heidelberg, 2007). 



\bibitem{fiedler1}
M. Fiedler. 
{\it Czechoslovak Mathematical Journal}  {\bf 23}, 298--305 (1973).

\bibitem{fiedler2}
M. Fiedler. 
{\it Czechoslovak Mathematical Journal} {\bf 25}, 619--633 (1975).

\bibitem{fiedler3}
M. Fiedler. 
{\it Combinatorics and Graph Theory} {\bf 25}, 57--70 (1989).

\bibitem{mohar}
B. Mohar.
in {\em Graph Theory, Combinatorics, and Applications}, Wiley publishers, 871-898 (1991).




\bibitem{clustering}
A. Y. Ng , M. I. Jordan and Y. Weiss.  
{\it Advances in Neural Information Processing Systems} (2001).

\bibitem{courant}
R. Courant.
{\it Math. Z.} {\bf 7} 1--57 (1920).

\bibitem{fisher}
E. Fischer. 
{\it Monatshefte f\"ur Math. und Phys.} {\bf 16}, 234--249 (1905).


\bibitem{kolmo}
V. Kolmogorov and R. Zabih. 
{\it IEEE T. Pattern Anal.}  {\bf 26}, 65--81 (2004).


\bibitem{land}
C. M. Reidys and P. F. Stadler. 
{\it SIAM Rev.} {\bf 44}, 3--54 (2002).

\bibitem{parisi}
M. M\'ezard, G. Parisi and M. A. Virasoro. 
{\it Spin glass theory and beyond}  
(World Scientific, Singapore, 1987). 

\bibitem{sale}
L. K. Grover. 
{\it Oper. Res. Lett.} {\bf 12}, 235--243 (1992).

\bibitem{phase}
S. J. Blundell and K. M. Blundell. 
{\it Concepts in Thermal Physics}. 
(Oxford University Press, Oxford, 2008).











\bibitem{jama}
A. Jamakovic and P. Van Mieghem. 
{\it NETWORKING'08 Proceedings of the 7th international IFIP-TC6 networking conference on AdHoc and sensor networks, wireless networks, next generation internet}, 183--194 (2008).





\end{thebibliography}
\end{document}